\documentclass[12pt]{article}
\usepackage{blindtext}
\usepackage{amsmath}
\usepackage{amssymb}
\usepackage{wrapfig}
\usepackage{multicol}
\usepackage[margin=1in]{geometry}
\usepackage{graphicx}
\usepackage{caption}
\usepackage{indentfirst}

\graphicspath{{Figures/}}
\emergencystretch\maxdimen
\renewenvironment{abstract}
{\par\noindent\textbf{\abstractname:}\ \ignorespaces}
{\par\medskip}

\makeatletter
\renewcommand\maketitle
{\noindent
{\large\bfseries\@title}%
\medskip\par
{\@author}%
\medskip\par
{\@date}%
\bigskip\par\noindent
}

\newenvironment{figurehere}
{\def\@captype{figure}}
{}

\renewcommand\section{\@startsection {section}{1}{\z@}%
{-1.5ex}%
{0.05ex }%
{\bfseries}}
\makeatother

\begin{document}
\onecolumn
\title{Euclidean Quantum Gravity from Variational Dynamics}
\author{Brenden McDearmon }
\maketitle

\begin{abstract}
A variational phase space is constructed for a compact and piecewise flat Riemannian manifold. An extended action functional is provided such that the variational dynamics generate a symplectic flow on the phase space. This symplectic flow is numerically integrated as it evolves with respect to the variational parameter. Assuming ergodicity, the resulting flow samples the Euclidean path integral.

\end{abstract}

\begin{multicols*}{2}
\section*{Introduction}

 In Euclidean quantum gravity, the causual structure of spacetime is omitted by replacing a pseudo-Riemannian metric of Lorentzian geometry with a positive definite metric of Riemannian geometry. Euclidean quantum gravity has been a useful tool for understanding aspects of relativistic quantum gravity such as cosmology, gravitational instantons, and blackhole thermodynamics \cite{gibbons1993euclidean} \cite{hartle1976path}\cite{gibbons1977action}\cite{hartle1983wave}. Euclidean quantum gravity has also proven useful in studying areas of mathematical interest including differential geometry and topology \cite{eguchi1979self}\cite{eguchi1978asymptotically}\cite{eguchi1980gravitation}. While Euclidean space can be Wick rotated to Minkowski space and reconstruction theorems are known in this specialized context, a more general reconstruction of relativistic quantum gravity from Euclidean quantum gravity has not been established \cite{osterwalder1973axioms}\cite{osterwalder1975axioms}\cite{glimm2012quantum}. Nevertheless, partial reconstructions and specialized applications are known \cite{baldazzi2019wicked}\cite{visser2017wick}\cite{ashtekar1999osterwalder}\cite{ashtekar2000constructing}\cite{lang2018hamiltonian}.

In typical Euclidean quantum gravity calculations, Monte Carlo methods or Langevin equations are used to generate field configurations that sample the Euclidean path integral \cite{hamber1986simplicial}\cite{hamber1994simplical}\cite{hartle1985simplicial}\cite{hamber1985nonperturbative}\cite{berg1985exploratory}\cite{loll1998discrete}. In a companion article directed to matter fields on Euclidean space, an alternative approach to sampling the Euclidean path integral, dubbed the ``variational dynamics,'' is provided \cite{mcdearmon2023euclidean}. In the variational dynamics approach, fields evolve with respect to a variational parameter $\lambda$ according to a symplectic flow on an extended phase space. Assuming ergodicity, the resulting flow samples the Euclidean path integral. Here, this technique is extended to include variational dynamics of a Riemannian manifold.

\section*{Variational Dynamics}

In variational dynamics on a d-dimensional Riemannian manifold $\mathcal{M}$, fields, such as the components of the metric tensor $ g_{ij}: x \in \mathcal{M} \mapsto \mathbb{R}$, are developed along the variational parameter $\lambda \in [0,\infty)$. To facilitate the variational development, the field is provided with a variationally conjugate field;  $\pi^{(g)}_{ij}$. Together with a global scalar, $s$, and its variational conjugate, $\pi^{(s)}$, the collection of degrees of freedom defines the variational phase space $\Gamma$. For example, a generic point in the variational phase space is given by $ \{\pi^{(g)}_{ij}, g_{ij},\pi^{(s)}, s\} \in \Gamma$ \footnote{Since this article is directed to the computational study of discretized systems, the exact nature of $\Gamma$ in the continuum is not precisely defined. A point in the discretized $\Gamma$ is given by $\{\times_{a\in A} \, \pi^{(g)}_{ij}(a), \times_{a\in A}\, g_{ij}(a), \pi^{(s)}, s\}$ where a indexes the discretized field and the index set A is finite.}. Given an action functional, $S : \gamma \in \Gamma \mapsto\mathbb{R}$, the variational dynamics is then defined by a set of variational equations such that the phase space trajectory, $\gamma (\lambda) \in \Gamma$, evolves by symplectomorphisms. Using the exemplary system of fields described above, the variational equations are the following.

 \begin{equation}\dot{\pi}^{(g)}_{ij}(\lambda) = \frac{d \pi^{(g)}_{ij}(\lambda+ \epsilon)}{d \epsilon} \vert_{\epsilon=0} = -\frac{\delta S}{\delta g_{ij}}(\lambda) \end{equation}
 \begin{equation}\dot{g}_{ij}(\lambda) = \frac{d g_{ij}(\lambda+ \epsilon)}{d \epsilon} \vert_{\epsilon=0} = \frac{\delta S}{\delta \pi^{(g)}_{ij}}(\lambda) \end{equation}
 \begin{equation} \dot{\pi}^{(s)}(\lambda) = \frac{d \pi^{(s)}(\lambda+ \epsilon)}{d \epsilon} \vert_{\epsilon=0} = -\frac{\partial S}{\partial s}(\lambda) \end{equation}
 \begin{equation} \dot{s}(\lambda) = \frac{d s(\lambda+ \epsilon)}{d \epsilon} \vert_{\epsilon=0} = \frac{\partial S}{\partial \pi^{(s)}}(\lambda) \end{equation}

While many action functionals may be defined, those studied herein are of the form $S=s \left( S^x-S^0 \right)$ where $S^x$ and $S^0$ are defined as follows.
\small
\begin{equation}
\begin{aligned} 
 S^{x}=  & \, S^v\left[\frac{\pi^{(g)}_{ij}}{s}, g_{ij}\right]+\frac{(\pi^{(s)})^2}{2m_s} +\hbar\, n_f\, ln(s)\\ &+S^g[g_{ij}] +S^{log}[g_{ij}]
\end{aligned} 
\end{equation} 

\begin{equation}
\begin{aligned}
 S^0= S^x \vert_{\lambda=0}
\end{aligned}
\end{equation} 
\normalsize

Here, $S^v\left[\frac{\pi^{(g)}_{ij}}{s}, g_{ij}\right]$ is a variational action, $S^g[g_{ij}]$ is a gravitational action, and $S^{log}[g_{ij}]$ is a functional of the metric which will play a role in determining the path integral measure as discussed below. Because only discretized systems will be considered in this article, $n_f$ is the finite number of degrees of freedom for the discretized fields which, in the systems studied herein, is equal to either the number of points (see, Example 1) or the number of 1-simplices (see, Examples 2-4)  in the discretized Remaniann manifold as discussed below. 

\section*{Discretization and Integration}

Discrete differential geometry is a useful tool in computational studies of manifolds \cite{crane2018discrete}. Various discretization techniques are known and have been applied to problems in mathematics and physics including quantum gravity and geometric flows \cite{regge1961general}\cite{regge2000discrete}\cite{loll1998discrete}\cite{miller2014simplicial}\cite{miller2014equivalence}\cite{hamber2010gravitational}\cite{kondo1984euclidean}. The discretization scheme used herein is the piecewise flat approximation advanced by Tullio Regge \cite{regge1961general}. In this approach, a Riemannian manifold is approximated as a simplicial complex. The simplicial complex consists of flat Euclidean simplices, $\sigma$, ranging in dimension from 0 (i.e. points) up to the dimension of the Reminanian manifold. The Riemannian metric is encoded in the set of 1-simplices (i.e. lines), $l_{ij}$, which measure the distance between end points. This set of distances can then be used to calculate the volumes and angles that define the geometry of the simplicial complex. 

For example, let $\mathcal{M}$ denote a discretized d-dimensional Riemannian manifold. The volume of $\mathcal{M}$ is calculated as the sum of volumes of the d-dimensional simplices contained in $\mathcal{M}$; $Vol(\mathcal{M}) = \sum_{\sigma^{(d)} \in \mathcal{M}} Vol^{(d)}(\sigma^{(d)})$.  The d-volume of a d-simplex, $Vol^{(d)}(\sigma^{(d)})$, can be calculated from the lengths of the 1-simplices which it contains \cite{veljan1995distance}\cite{hartle1985simplicial}. Together with the (d-2)-volumes of the codimension 2 simplices, curvature terms depend on the defect angles occurring at the codimension 2 simplices defined as $ \theta(\sigma^{(d-2)}) = 2 \pi - \sum_{\sigma^{(d)} \supset \sigma^{(d-2)}} \theta^{DH}(\sigma^{(d-2)}, \sigma^{(d)})$. Here, $ \theta^{DH}(\sigma^{(d-2)}, \sigma^{(d)})$ is the dihedral angle in $\sigma^{(d)}$ at $\sigma^{(d-2)}$ which can be calculated from the lengths of the 1-simplices contained in $\sigma^{(d)}$ \cite{hartle1985simplicial}. For example, the integral of the scalar curvature over the manifold is given by the sum over each (d-2)-simplex of the defect angle times the (d-2)-volume; $ R(\mathcal{M})= \int_{\mathcal{M}} R \, dvol = 2 \sum_{\sigma^{(d-2)}} \theta(\sigma^{(d-2)}) Vol^{(d-2)}(\sigma^{(d-2)})$ \cite{hartle1985simplicial}\cite{regge1961general}. Formulas for higher order curvature terms in simplicial complexes have also been developed by Hamber and Willams \cite{hamber1986simplicial}\cite{hamber1984higher}. For example, the integral of the  Riemann curvature squared term over the manifold is given by the following sum $Rm^2(\mathcal{M})=\int_{\mathcal{M}} Rm^2 \, dvol = 4 \sum_{\sigma^{(d-2)}} \theta(\sigma^{(d-2)})^2 Vol^{(d-2)}(\sigma^{(d-2)})^2$  \cite{hamber1986simplicial}\cite{hamber1984higher}. The convergence of various discretized curvature terms in the continuum limit has been studied \cite{cheeger1984curvature}.

In the variational dynamics equations, derivatives of simplicial volumes and dihedral angles with respect to edge-length may occur. Hartle provided a convenient formula for calculating the derivative of the d-volume of an d-dimensional simplex with respect to edge-length; $\frac{\partial Vol^{(d)}(\sigma^{(d)})}{\partial l_{ij}}$ \cite{hartle1985simplicial}. The derivative of a dihedral angle in a d-simplex with respect to edge-length, $\frac{ \partial \theta^{DH}((\sigma^{(d-2)} , (\sigma^{(d)}))}{l_{ij}}$, can be calculated using a formula provided by Dittrich, Freidel, and Speziale \cite{dittrich2007linearized}; see, also, \cite{dittrich2012path}. Of particular note, the derivative of the integrated scalar curvature term with respect to edge length, $\frac{\partial R(\mathcal{M})}{\partial l_{ij}}$, is given by $\sum_{\sigma^{(d-2)}} \theta(\sigma^{(d-2)}) \frac{ \partial Vol^{(d-2)}(\sigma^{(d-2)})}{\partial l_{ij}}$ because of Schlaffli's identity \cite{regge1961general}\cite{milnor1994schlafli}\cite{rivin2000schlafli}. 

Once the metric has been discretized, a $\lambda=0$ configuration is initialized. That is, a point $\gamma(0) \in \Gamma$ in the variational phase space is specified.
\small
\begin{equation}
\begin{aligned}
\gamma(0)=\{\times_{(i,j)} \pi^{(l)}_{ij}(0), \times_{(i,j)} l_{ij}(0), \pi^{(s)}(0), s(0)\}
\end{aligned}
\end{equation} 
\normalsize

 The trajectory of this point as the system evolves with respect to the variational parameter $\lambda$ is then obtained by numerically integrating the discretized variational equations. Because the discretized variational equations are not in general separable, the generalized leap-frog algorithm used in the companion article \cite{mcdearmon2023euclidean} is not applicable. For this reason, the semiexplicit symplectic integrator developed by Jayawardana and Ohsawa was used \cite{jayawardana2023semiexplicit}. 

\section*{The Euclidean Path Integral}

Depending on the specifics of the action functional, the variational dynamics may or may not exist for all $\lambda \in [0,\infty)$. Nonetheless, assuming that the variational dynamics does exist for all $\lambda \in [0,\infty)$, and further assuming that the flow is ergodic, the variational dynamics can be shown to sample the Euclidean path integral. 

By the variational equations (1) to (4), $\frac{dS}{d\lambda}=0$ so that the extended action functional is conserved by the variational dynamics. Further, by definition $S|_{\lambda=0}=0$. Accordingly, the partition function for the system is defined by equation (8). 

\begin{equation}
\begin{aligned} 
\mathcal{Z} = \int \delta \left( s \left(S^x-S^0 \right)\right) d\Gamma
\end{aligned}
\end{equation}

The phase space measure $d\Gamma$ is defined as follows:
\begin{equation}
\begin{aligned} 
 d\Gamma =ds\,d\pi^{(s)}\,D\left[\pi^{(l)}\right]\,D\left[l\right].
\end{aligned}
\end{equation}

Here, $D[l]=\prod_{(i,j)} dl_{ij}$ and $D\left[\pi^{(l)}\right]=\prod_{(i,j)} d\pi^{(l)}_{ij}$. Now, define a change of variables $\tilde{\pi}^{(l)}_{ij}=\frac{\pi^{(l)}_{ij}}{s}$, and notice that $d\pi^{(g)}_{ij} =s d \tilde{\pi}^{(g)}_{ij}$ for s held constant. This change of variables thus results in a change of measure.
\begin{equation}
\begin{aligned} 
 d\tilde{\Gamma} =s^{n_f} ds\,d\pi^{(s)}\,D\left[\tilde{\pi}^{(l)}\right]\,D\left[l\right]
\end{aligned}
\end{equation}

With this change of variables, integration over the Dirac delta function with respect to $s$ can be performed using the identity $ \frac{ d }{ds} \delta\left[f(s)\right]= \frac{\delta\left[ f(s-s')\right]}{\left( \frac{df}{ds}|_{(s')}\right)}$, where $s'$ is the isolated zero of $S$ with respect to $s$. 

\scriptsize
\begin{equation}
\begin{aligned} 
&s' = \\& exp  \big[  -\big(S^v[\tilde{\pi}^{(l)}, l]+\frac{(\pi^{(s)})^2}{2m_s}+ S^g[l]+S^{log}[l]-S^0 \big) / ( \hbar n_f )  \big]
\end{aligned}
\end{equation}
\normalsize

Evaluating the integral with respect to s gives the following equation where $C$ is a constant.

\scriptsize
\begin{equation}
\begin{aligned} 
\mathcal{Z} =&C \int d\pi^{(s)}\,D\left[\tilde{\pi}^{(l)}\right]\,D\left[l\right] \\& exp  \big[ -\big(S^v[\tilde{\pi}^{(l)}, l]+\frac{(\pi^{(s)})^2}{2m_s}+S^g[l]+S^{log}[l]-S^0 \big) / \hbar  \big] 
\end{aligned}
\end{equation}
\normalsize

 Next, while more general variational actions are permissible, from here on, the variational action will be restricted to the form $S^v[\tilde{\pi}^{(l)}, l]= \sum_{(i,j)} \,k_l[l]\frac{\left(\tilde{\pi}^{(l)}_{ij}\right)^2}{2}$. Here, $k_l[l]$ depends on the set of distances $l_{ij}$. Because the integrals with respect to $d\pi^{(s)}$ and $D\left[\tilde{\pi}^{(l)}\right]$ are Gaussian, they can be evaluated to give the following equation where $C'$ is a new constant.
\scriptsize
\begin{equation}
\begin{aligned} 
& \mathcal{Z} = \\& C' \int \left( \prod_{l} \frac{1}{\sqrt{k_l[l]}} \right)  D\left[l\right]  exp \left [ \left.  -\left(S^g[l]+S^{log}[l]-S^0 \right) \middle / \hbar \right. \right]
\end{aligned}
\end{equation}
\normalsize

Further restricting to $S^{log}[l]=\sum_{l} log\left(a_l[l]\right)$, equation (14) provides the Euclidean path integral.

\begin{equation} 
 \mathcal{Z} = Z^0 \int exp \left [  -S^g[l]/ \hbar \right] F[l] D\left[l\right]
\end{equation}

Here, $Z^0$ is a constant and $F[l]=\left( \prod_{l} \frac{1}{a_l[l] \sqrt{k_l[l]}} \right) $ is a component of the path integral measure. Various such $F[l]$ have been studied in the context of Euclidean quantum gravity \cite{hamber1999measure}\cite{dittrich2012path}\cite{dittrich2014discretization}. Thus, assuming ergodicity, the partition function for the system undergoing variational dynamics is the Euclidean path integral.

\section*{Expectation Values}

The sequence of field configurations generated by the variational dynamics can be used to calculate expectation values of observables \footnote{Here,``observables'' refer to Euclidean observables. Schrader provided a construction for quantum observables for a particular simplicial Euclidean path integral \cite{schrader2016reflection}.}.

For example, for some observable $\mathcal O \left[l\right]$ taken to be a functional of the metric, the expectation value with respect to the variational flow is given by the following.

\begin{equation} 
\left<\mathcal O \left[l\right]\right>_{\lambda}= \lim_{\Lambda \to \infty}\frac{1}{\Lambda} \int_0^{\Lambda} \mathcal O \left[l(\lambda)\right] d \lambda
\end{equation}

The expectation value of the observable  $\mathcal O \left[l\right]$ averaged over the variational phase space is given by the following.

\begin{equation}
\begin{aligned} 
&\left<\mathcal O \left[l\right]\right>_{\Gamma}=\\& \frac{  \int \mathcal O \left[l\right] exp \left [  -\left(S^g[l] \right) / \hbar \right] F[l] D\left[l\right]}{\int exp \left [  -\left(S^g[l] \right) / \hbar \right] F[l] D\left[l\right]}
\end{aligned}
\end{equation}

 By the assumption of ergodicity, $ \left<\mathcal O \left[l\right]\right>_{\lambda}= \left<\mathcal O \left[l\right]\right>_{\Gamma}$.

\section*{Examples}

In each of the following examples, the manifold was initialized with a flat geometry. While other boundary conditions are possible \cite{hartle1981boundary}, each of the examples presented herein used periodic boundary conditions. The initial manifold was discretized as a regular periodic lattice of $N=4 \times 4 \times 4 \times 4$ points having a lattice spacing of 1 with each unit cell of the lattice having 24 4-simplices.   Further, $s$ was initialized at 1, $\hbar$ was set equal to 1, and $m_s$ was set equal to $0.5 N$. Other than $\pi^{(s)}$ which was initialized at 0, the remaining variational fields were initialized by randomly selecting each value from a uniform distribution extending symmetrically around 0 with a range such that $\frac{2 S^{v} \vert_{\lambda=0}}{n_f \hbar}$ was approximately equal to 1.

Once an initial configuration was established, each example was numerically integrated with respect to $\lambda$ using Jayawardana and Ohsawa's semiexplicit symplectic integration method with a step size of $d\lambda=0.0001$ and a tolerance of 0.0001. The systems studied in the examples were stable for the number of steps observed. This stability assessment is based on the apparent boundedness of various parameters (e.g., the manifold volume $Vol\left(\mathcal{M}\right)$ and the error term $ \frac{S^x-S^0}{S^0}$) as well as a non-collapsing criteria whereby all n-simplicies in the d-dimensional manifold of each dimension n ranging from 1 to d had a positive non-zero n-volume for each step during the simulation. The results provided in the examples are not intended to be quantitatively accurate or predictive, and the action functionals used are not intended to model a physical system. Rather, the results demonstrate the proof-of-concept use of variational dynamics for Euclidean quantum gravity calculations. Improved quantitative results could be provided by, e.g., increasing the number of points in the lattice, decreasing the step size $d \lambda$, increasing the total number of steps, and, if necessary, including an ``equilibration'' procedure.

\subsection*{\small \;\;1.\;Conformal Variations}

As a first example, instead of including all variations of the discretized metric tensor, only conformal variations were considered. Given a reference metric $l^{(0)}_{ij}$, a discrete conformal transformation is given by $l_{ij} = l^{(0)}_{ij}e^{f_i+f_j}$ where $f_p$ is a real-scalar field evaluated at point $p$ \cite{glickenstein2011discrete}\cite{crane2019conformal}\cite{rovcek1984quantization}. Because the reference metric in this example corresponds to a flat geometry, the metrics generated by the variational dynamics are all conformally flat.  In the context of quantum gravity, conformal deformations are interesting to study because the Einstein-Hilbert action, $R\left( \mathcal{M}\right)=2 \sum_{\sigma^{(2)}} \left(Vol^{(2)}(\sigma^{(2)})\theta(\sigma^{(2)})\right)$, has a conformal instability and is not bounded from below \cite{gibbons1978path}. By restricting to conformal deformations and modifying the gravitational action to include additional terms depending on the conformal field $f$, one can arrive at a stable system that undergoes metric variations within a conformal class. More broadly, optimization problems over conformal deformations of a metric have also been applied to various problems of mathematical interest such as the Yambe problem \cite{glickenstein2011discrete}, machine learning \cite{maron2017convolutional}, and other problems in mathematical physics \cite{york1972role}. 

The variational dynamics of this example was defined by the following action.

\scriptsize
\begin{equation}
\begin{aligned} 
S= s \Bigg( &\sum_p \left(\frac{\left(\pi^{(f)}\right)^2}{2s^2}e^{-4f_p} V_p-\frac{1}{2} log(V_p) \right) \\&+R\left( \mathcal{M}\right)+\sum_p \left(3\Delta(f)\vert_p V_p+4\left(f_p-\frac{1}{4}\right)V_p\right)  \\&+\sum_{(i,j)}3\left(\frac{f_i-f_j}{l_{ij}}\right)^2 V_l+\frac{\left(\pi^{(s)}\right)^2}{2N}+N log(s)-S^0 \Bigg)
\end{aligned}
\end{equation}
\normalsize

Here, $V_p=\sum_{\sigma^{(4)} \supset p}\frac{Vol^{(4)}(\sigma^{(4)})}{5}$ is the 4-volume assigned to point $p$ and is the volume of the 4-simplices $\sigma^{(4)}$ which contain $p$ divided by the number of points in a 4-simplex (i.e. 5). Similarly, $V_l=\sum_{\sigma^{(4)} \supset l}\frac{Vol^{(4)}(\sigma^{(4)})}{10}$ is the 4-volume assigned to the 1-simplex $l$ and is the volume of the 4-simplices $\sigma^{(4)}$ which contain $l$  divided by the number of 1-simplices in a 4-simplex (i.e. 10). The positive-semidefinite laplacian of $f$ evaluated at $p$ is defined as $\Delta(f)\vert_p=\sum_{l_{ip}} \frac{f_p-f_i}{l_{ip}^2}$. In this system, $m_s=N$ and the number of discretized degrees of freedom was $n_f=N$. 

After initializing the system in accordance with the discussion above, the resulting variational dynamics was stable over the number of steps computed as shown in Figure 1. In particular, Figure 1C shows that the magnitude of the total error, $\vert \frac{S^x-S^0}{S^0}\vert \, 100 \%$, was stably bounded at below 0.001 \% over the 500,000 steps computed. Similarly, Figure 1 demonstrates that various components of the action stably fluctuated as the system evolved with respect to $\lambda$. For example, the gravitational action $S^g$ was observed to fluctuate about a mean $\left<\frac{S^g}{N}\right>=-0.81$ with a variance $\left<\left(\frac{S^g}{N}-\left<\frac{S^g}{N}\right>\right)^2\right>=0.074$ [Figure 1D]. 

Interestingly, the volume of the manifold was observed to oscillate between about $\frac{Vol\left(\mathcal{M}\right)}{Vol\left(\mathcal{M}\right)\vert_{\lambda=0}}=4.1$ and 0.97 [Figure 1F].

\end{multicols*}
\begin{figure*}[h]
\centering
\includegraphics[width=0.85\textwidth]{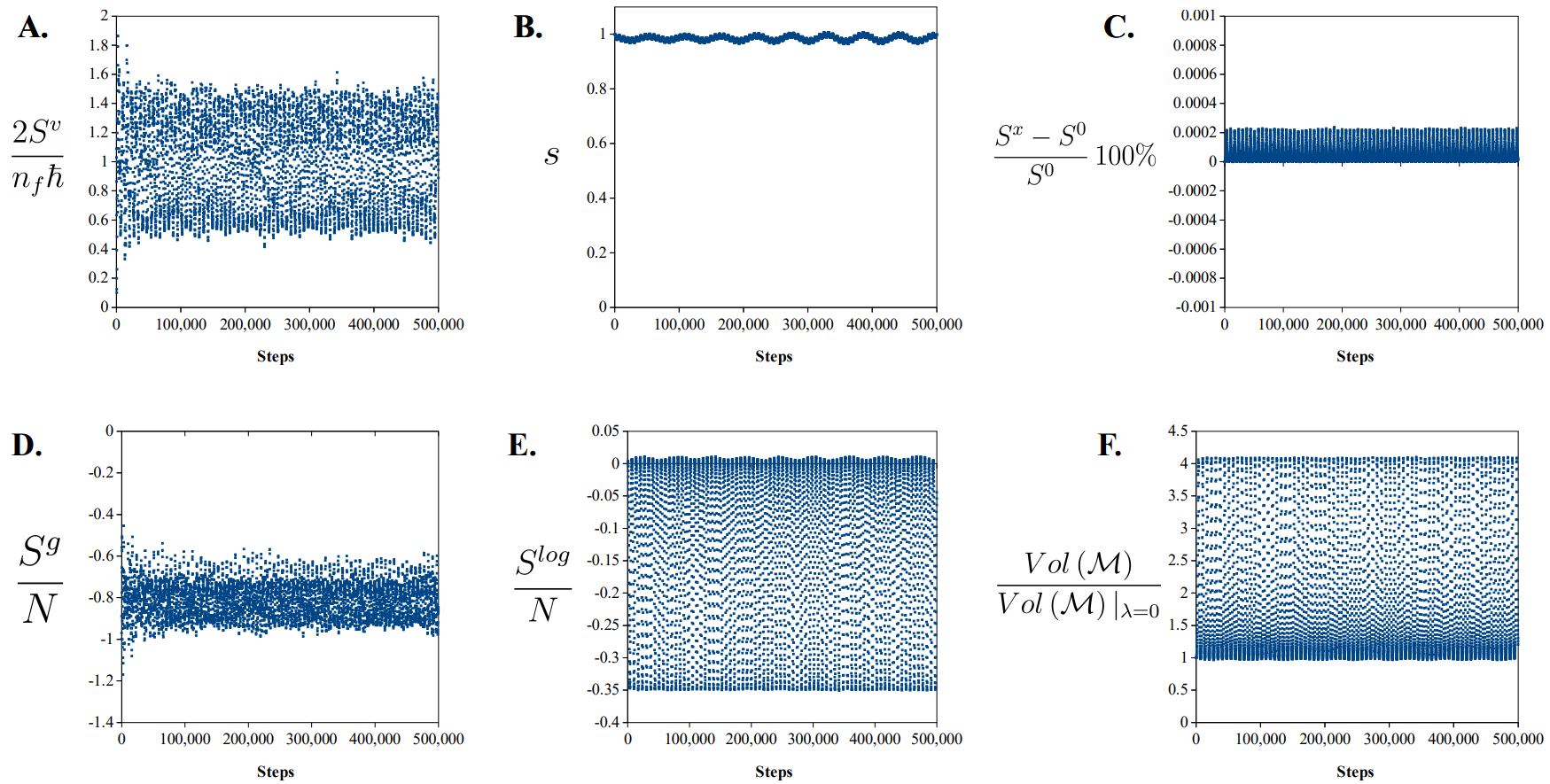}
\caption{Properties of the system according to Example 1 as it undergoes conformal variations.}
\end{figure*}
\begin{multicols*}{2}
 Figure 2 depicts the integrated scalar curvature and Riemann curvature of the manifold. The manifold was observed to have a small positive scalar curvature on average with $\left< \frac{R(\mathcal{M})}{N}\right>=0.24$ [Figure 2A]. Together with the larger average integrated Riemann curvature $\left<\frac{ Rm^2(\mathcal{M})}{N}\right>=106$ observed in Figure 2B, this indicates a broader distribution of local curvature between both positive and negative values. 

This data demonstrates possibly stable variational dynamics for a manifold undergoing conformal deformations. Assuming ergodicity, the Euclidean path integral for this system is given by the following:

\scriptsize
\begin{equation}
\begin{aligned} 
\mathcal{Z} &= Z^0 \int  exp  \Bigg[-\Bigg(R\left( \mathcal{M}\right)+\sum_{(i,j)}3\left(\frac{f_i-f_j}{l_{ij}}\right)^2 V_l \\& +\sum_p \left(3\Delta(f)\vert_p V_p+4\left(f_p-\frac{1}{4}\right)V_p\right)\Bigg) \Bigg] D\left[e^{-2f}\right].
\end{aligned}
\end{equation}
\normalsize

\begin{figurehere}
\centering
\includegraphics[width=0.85\linewidth]{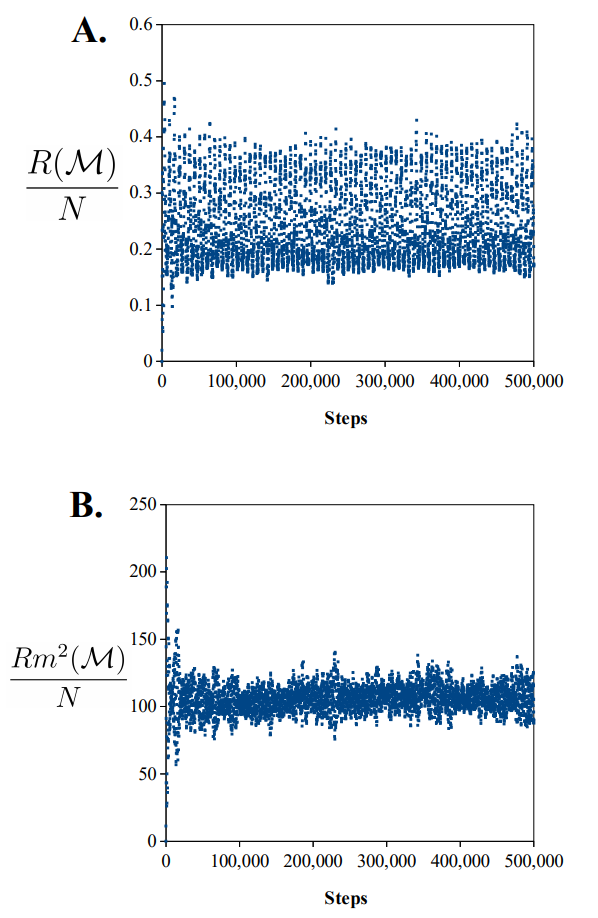}
\caption{The total scalar curvature (A) and total Riemannian curvature square (B) integrated over $\mathcal{M}$.}
\end{figurehere}

\end{multicols*}
\begin{figure*}[h]
\centering
\includegraphics[width=0.95\textwidth]{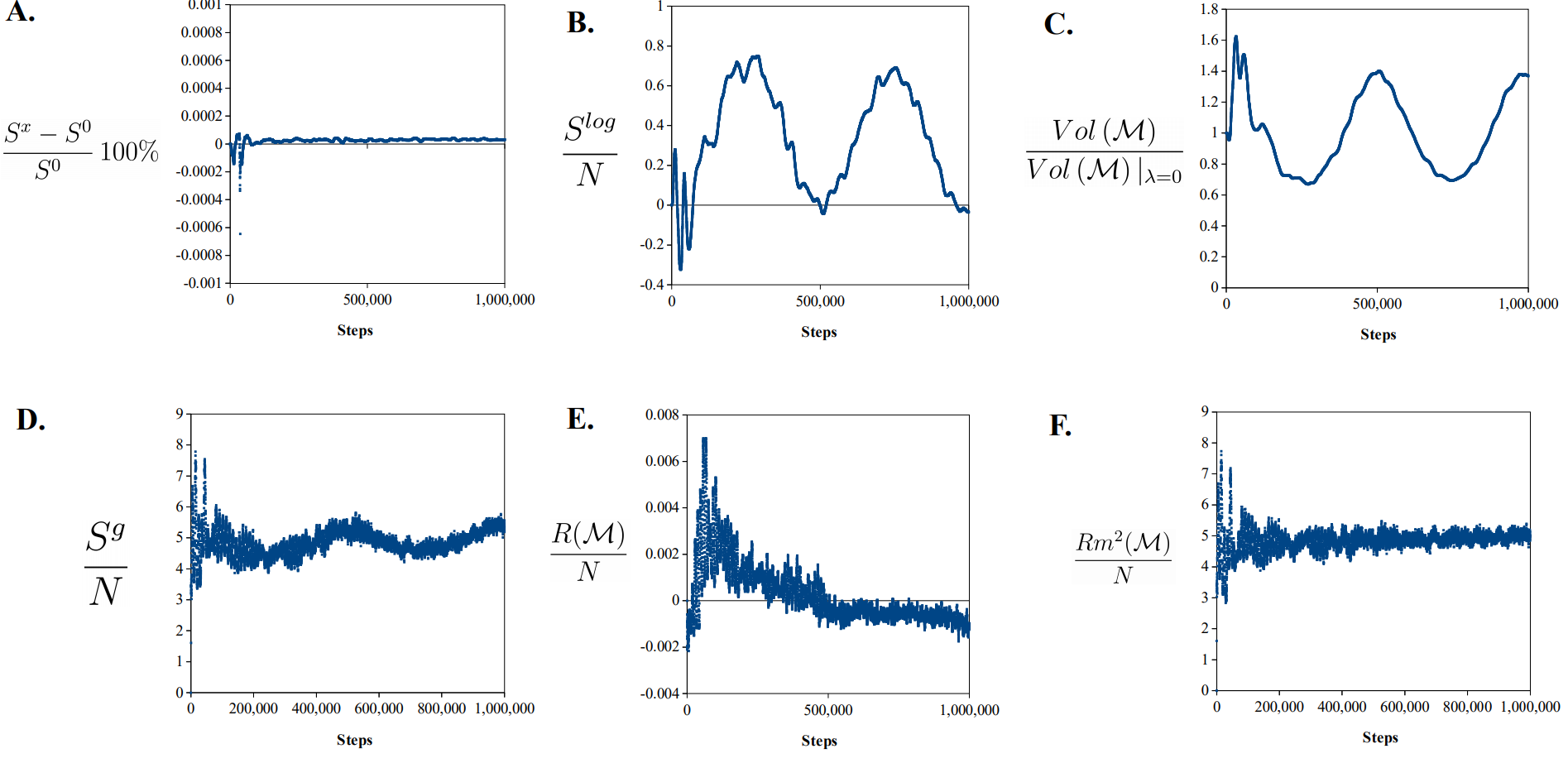}
\caption{Properties of the system according to Example 2.}
\end{figure*}
\begin{multicols*}{2}

\subsection*{\small \;\;2.\; Cosmological Constant}
Continuing on to variations of the full discrete metric tensor, various stable systems can be prescribed. In order to have a gravitational action that is bounded from below, it is helpful to include higher-order curvature terms \cite{hamber1986simplicial}\cite{hamber1984higher}. The Riemann curvature squared term, $Rm^2(\mathcal{M})= 4 \sum_{\sigma^{(d-2)}} \theta(\sigma^{(d-2)})^2 Vol^{(d-2)}(\sigma^{(d-2)})^2$, is the easiest such term to compute. One additional difficulty encountered when using gravitational actions that contain an Einstein-Hilbert term is that the scalar curvature is not scale invariant. Thus, the total manifold volume may have a tendency to diverge when undergoing variational dynamics. In Monte Carlo simulations of similar systems, this volume divergence can be avoided by including a cosmological constant \cite{hamber1999measure}\cite{loll1998discrete}. With these considerations in mind, variational dynamics of a system defined by the following action was studied.
\scriptsize
\begin{equation}
\begin{aligned} 
S= s \Bigg( &\sum_{(i,j)} \left(\frac{\left(\pi^{(l)}_{ij}\right)^2}{2s^2\left(l_{ij}\right)^2} V_l-\frac{1}{15} log(V_l) -\frac{1}{2} log(w_l) \right)\\&+R(\mathcal{M})+Rm^2(\mathcal{M})+ Vol(\mathcal{M})\\&+\frac{\left(\pi^{(s)}\right)^2}{2N}+15N log(s)-S^0 \Bigg)
\end{aligned}
\end{equation}
\normalsize
Here, $w_l=\frac{\left(\prod_{\sigma^{(4)} \supset l}Vol\left(\sigma^{(4)} \right)\right)^{\frac{1}{n_l}}}{V_l}$ is proportional to the ratio of the geometric mean volume at $l$ and the arithmetic mean volume at $l$. Interestingly,the geometric mean divided by arithmetic mean ranges from 0 to 1 and equals 1 only when the volumes of the 4-simplices containing $l$ are equal \cite{uchida2008simple}. Thus, modulo a possible non-dynamical constant, $\sum_{(i,j)}\left( -\frac{1}{2} log(w_l)\right) \geq 0$ with equality holding only when all of the 4-simplices in $\mathcal{M}$ have equal volume. Further, the term $\sum_{(i,j)}\left( -\frac{1}{2} log(w_l)\right) \rightarrow \infty$ when the volume of any 4-simplex in $\mathcal{M}$  goes to 0. This term thus measures the distribution of simplicial 4-volumes in the manifold and helps prevent local volume collapsing.

Figure 3 demonstrates that this system was stable over the 1,000,000 steps observed. Again, the magnitude of the error term was less than 0.001\% [Figure 3A]. There appears to be a transient period of about 100,000 steps after which the manifold's volume adopts an oscillatory mode [Figure 3C]. 

\end{multicols*}
\begin{figure*}[h]
\centering
\includegraphics[width=0.95\textwidth]{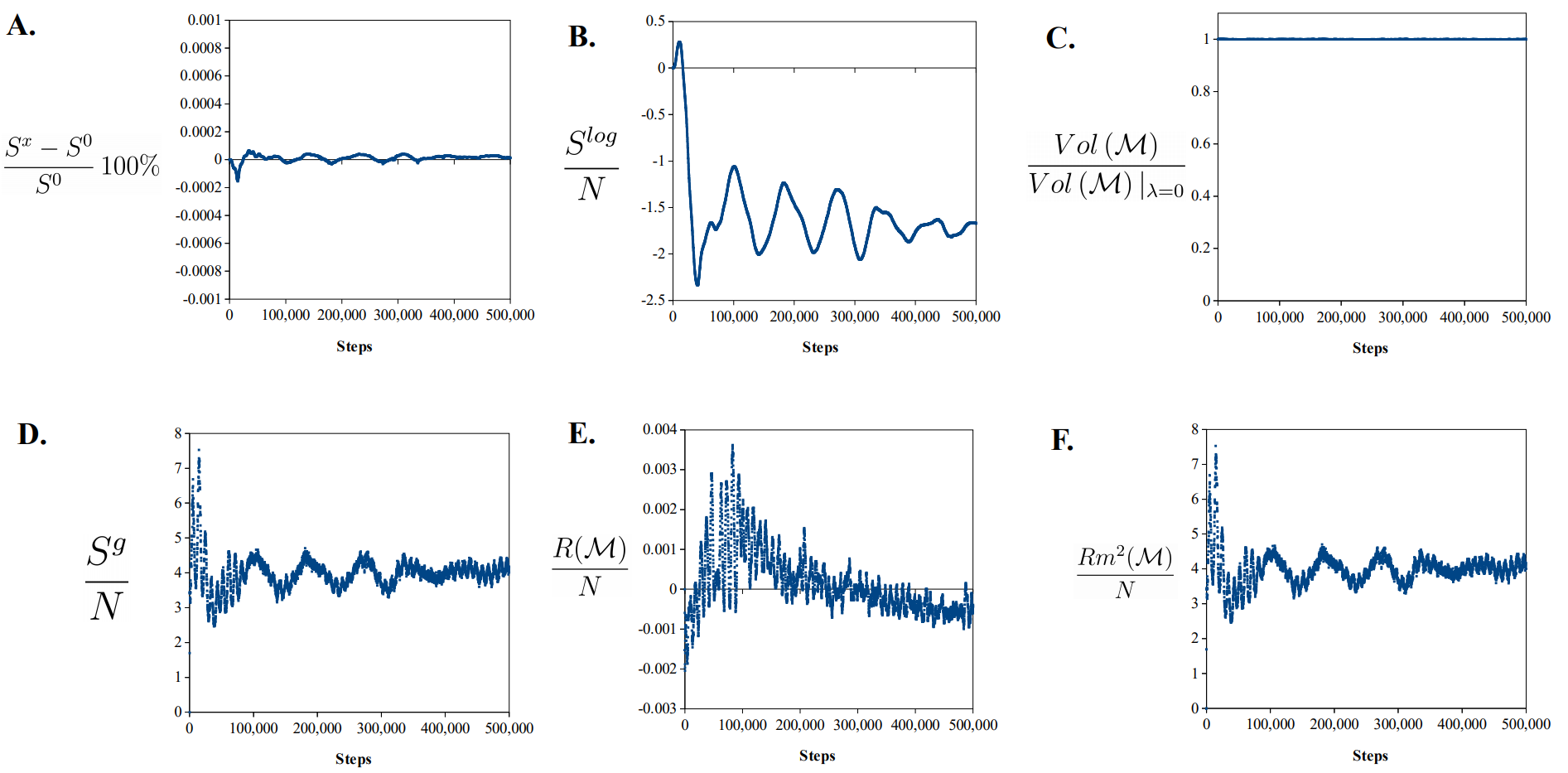}
\caption{Properties of the system according to Example 3 with volume fixing term.}
\end{figure*}
\begin{multicols*}{2}

The integrated scalar curvature  was small taking on both positive and negative values, and the integrated Riemannian curvature was considerably smaller than in Example 1 [Figures 3E-F]. Example 2 thus demonstrates a possibly stable variational dynamics for the full discretized metric tensor in a system with a higher-order gravity action including a cosmological constant. Assuming ergodicity, the Euclidean path integral for this system is given by the following:

\scriptsize
\begin{equation}
\begin{aligned} 
\mathcal{Z} = Z^0 \int  &exp  \Bigg[-\Bigg( R(\mathcal{M})+Rm^2(\mathcal{M})+ Vol(\mathcal{M})\Bigg) \Bigg] \\& \prod_{l}\left( \frac{w_l^{\frac{1}{2}}}{V_l^{\frac{13}{30}}}\right)D\left[l^2\right]. 
\end{aligned}
\end{equation}
\normalsize

\subsection*{\small \;\;3.\; Volume Fixing}

Another technique sometimes used in Monte Carlo simulations of Euclidean quantum gravity systems for keeping the manifold volume from diverging is to continuously re-scale the manifold to a fixed volume \cite{berg1985exploratory}\cite{loll1998discrete}. An alternative approach used here is to include a volume fixing term such as $f(N)\left(Vol(\mathcal{M})-Vol(\mathcal{M})\vert_{\lambda=0}\right)^2$ where $f(N)$ is an increasing function of N. Such a volume fixing term was incorporated into the following action:

\scriptsize
\begin{equation}
\begin{aligned} 
S= s \Bigg( &\sum_{(i,j)} \left(\frac{\left(\pi^{(l)}_{ij}\right)^2}{2s^2\left(l_{ij}\right)^2} V_l-\frac{1}{2} log\left(\frac{V_l}{\left(l_{ij}\right)^4}\right) -\frac{1}{2} log(w_l) \right)\\&+Rm^2(\mathcal{M})+0.75N\left(Vol(\mathcal{M})-Vol(\mathcal{M})\vert_{\lambda=0}\right)^2\\&+\frac{\left(\pi^{(s)}\right)^2}{2N}+15N log(s)-S^0 \Bigg). 
\end{aligned}
\end{equation}
\normalsize

This system resulted in a stable variational dynamics, and the magnitude of the error term was less than 0.001\% [Figure 4A]. As seen in Figure 4C, the volume of the manifold was relatively constant during the variational dynamics with $\left<\frac{Vol\left(\mathcal{M}\right)}{Vol\left(\mathcal{M}\right)\vert_{\lambda=0}}\right>=1.0$ with a variance of $3.7 \cdot 10^{-8}$. Assuming ergodicity, the Euclidean path integral for this system is given by the following:

\scriptsize
\begin{equation}
\begin{aligned} 
&\mathcal{Z} =Z^0 \int  \prod_{l}\left( \frac{\sqrt{w_l}}{l}\right) D\left[l\right]  \\& exp  \Bigg[-\Bigg( Rm^2(\mathcal{M})+ 0.75N\left(Vol(\mathcal{M})-Vol(\mathcal{M})\vert_{\lambda=0}\right)^2\Bigg) \Bigg].
\end{aligned}
\end{equation}
\normalsize

\end{multicols*}
\begin{figure*}[b]
\centering
\includegraphics[width=0.95\textwidth]{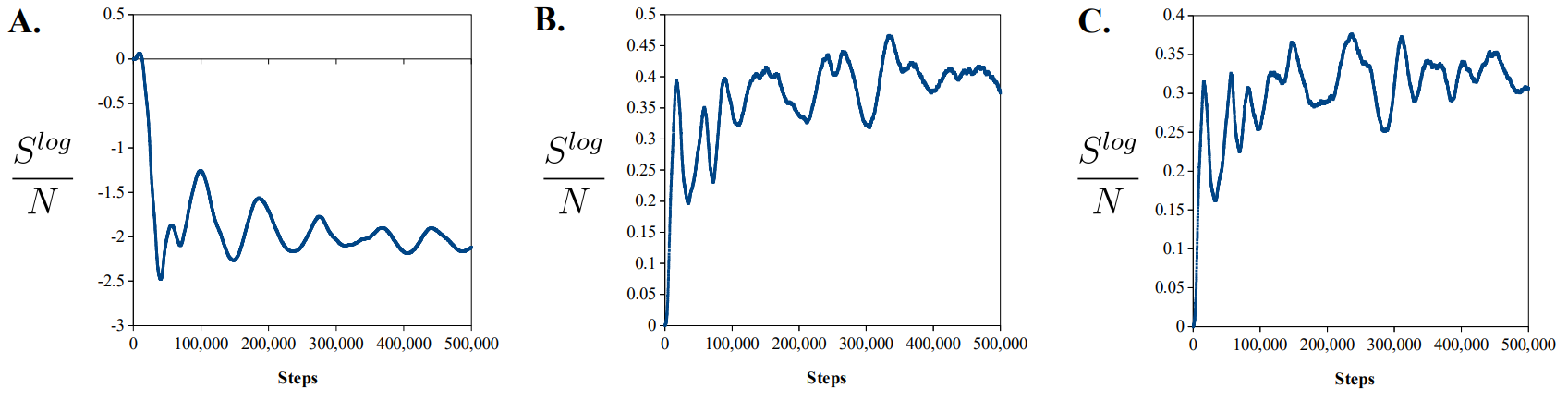}
\caption{Comparison of $S^{log}$ terms for the systems of Example 4 having differing expected functional integration measures.}
\end{figure*}
\begin{multicols*}{2}

Interestingly, apart from the volume fixing term $ 0.75N\left(Vol(\mathcal{M})-Vol(\mathcal{M})\vert_{\lambda=0}\right)^2$, the Euclidean path integral is scale invariant.  Moreover, with an appropriate constant $Z^0$, which can be suitably modified by adding a non-dynamical constant to the extended action $S^x$, the volume fixing term approaches a Dirac delta as $N \rightarrow \infty$. That is, equation (23) approaches the following when N is large and where $\tilde{Z}^0$ is a new constant.

 \scriptsize
\begin{equation}
\begin{aligned} 
\mathcal{Z} =& \tilde{Z}^0 \int   \delta\left(\left(Vol(\mathcal{M})-Vol(\mathcal{M})\vert_{\lambda=0}\right)^2\right) \\&  exp  \Bigg[-Rm^2(\mathcal{M}) \Bigg] \prod_{l}\left( \frac{\sqrt{w_l}}{l}\right) D\left[l\right]
\end{aligned}
\end{equation}
\normalsize

\subsection*{\small \;\;4.\; Functional Integration Measures}

Various integration measures have been proposed for the Euclidean path integral \cite{hamber1999measure}\cite{dittrich2012path}\cite{dittrich2014discretization}\cite{loll1998discrete}. An interesting aspect of the variational dynamics approach to the Euclidean path integral is that the measure can depend on, e.g., the coupling constants of the $k_l$ in the variational action $S^v$  as well as the action component $S^{log}$. In particular, the same integration measure may, assuming ergodicity, be expected to result from combinations of different $S^v$ and $S^{log}$; however, some such combinations may not produce a stable variational flow. To demonstrate systems expected to produce Euclidean path integrals with different integration measures, the system of Example 3 was repeated but with different $S^v$ and $S^{log}$ as follows.
 \scriptsize
$$\textbf{A.} \,S^v+S^{log}=\sum_{(i,j)} \left(\frac{\left(\pi^{(l)}_{ij}\right)^2}{2s^2\left(l_{ij}\right)^2} V_l -\frac{1}{2} log\left(\frac{V_l}{\left(l_{ij}\right)^4}\right) \right) $$
$$\textbf{B.} \,S^v+S^{log}=\sum_{(i,j)} \left(\frac{\left(\pi^{(l)}_{ij}\right)^2}{2s^2\left(l_{ij}\right)^2} w_l V_l -\frac{1}{2} log(w_l) \right) $$
$$\textbf{C.} \,S^v+S^{log}=\sum_{(i,j)} \left(\frac{\left(\pi^{(l)}_{ij}\right)^2}{2s^2\left(l_{ij}\right)^2} V_l -\frac{1}{2} log(w_l) \right) $$
\normalsize

Each of these systems resulted in a stable variational dynamics over the number of steps observed. Figure 5 depicts $\frac{S^{log}}{N}$ for each of these systems, and Figures A1-A3 of the appendix reports additional properties. Assuming ergodicity and using $ \delta\left(\left(Vol(\mathcal{M})-Vol(\mathcal{M})\vert_{\lambda=0}\right)^2\right)$ to represent the volume fixing term as in Example 3, the corresponding Euclidean path integrals are given by the following.
 \scriptsize
\begin{equation*}
\begin{aligned} 
\textbf{A.} \, \mathcal{Z} =& \tilde{Z}^0 \int   \delta\left(\left(Vol(\mathcal{M})-Vol(\mathcal{M})\vert_{\lambda=0}\right)^2\right) \\&  exp  \Bigg[-Rm^2(\mathcal{M}) \Bigg] \prod_{l}\left( \frac{1}{l}\right) D\left[l\right]
\end{aligned}
\end{equation*}
\normalsize

 \scriptsize
\begin{equation*}
\begin{aligned} 
\textbf{B.} \, \mathcal{Z} =& \tilde{Z}^0 \int   \delta\left(\left(Vol(\mathcal{M})-Vol(\mathcal{M})\vert_{\lambda=0}\right)^2\right) \\&  exp  \Bigg[-Rm^2(\mathcal{M}) \Bigg] \prod_{l}\left( \frac{l}{\sqrt{V_l}}\right) D\left[l\right]
\end{aligned}
\end{equation*}
\normalsize

 \scriptsize
\begin{equation*}
\begin{aligned} 
\textbf{C.}\,  \mathcal{Z} =& \tilde{Z}^0 \int   \delta\left(\left(Vol(\mathcal{M})-Vol(\mathcal{M})\vert_{\lambda=0}\right)^2\right) \\&  exp  \Bigg[-Rm^2(\mathcal{M}) \Bigg] \prod_{l}\left( \frac{l\sqrt{w_l}}{\sqrt{V_l}}\right) D\left[l\right]
\end{aligned}
\end{equation*}
\normalsize

As in Example 3, each of these Euclidean path integrals are scale invariant except for the volume fixing term. Together, Example 3 and Examples 4A-C demonstrate potentially stable variational dynamics with different $S^v$ and $S^{log}$ such that the expected Euclidean path integral measure is changed. Such examples may enable a systematic study of the influence of the path integral measure of Euclidean quantum gravity in the context of variational dynamics.

\section*{Discussion}

These results demonstrate the proof-of-concept use of variational dynamics for generating field configurations which, assuming ergodicity, sample the Euclidean path integral. As such, this provides a new method for performing Euclidean quantum gravity calculations. Verification of the accuracy, and assumed ergodicity, of this new technique should be conducted by, e.g., comparing these results to field configurations generated using traditional Monte Carlo methods. 

One advantage variational dynamics may have over Monte Carlo methods is the availability of new analytical techniques which make use of correlations with respect to the variational parameter. For example, one could consider various correlation functions with respect to $\lambda$ given by $C_{A,B}(\tilde{\lambda}) = lim_{\Lambda \to \infty}\int_0^{\Lambda}\mathcal O_A \left[l(\lambda-\tilde{\lambda})\right] \mathcal O_B \left[l(\lambda\right] d\lambda$. In statistical mechanics, analogous correlation functions, and their Fourier transforms, are used extensively to relate dynamical properties of a system to equilibrium properties \cite{zwanzig1965time}\cite{kubo1966fluctuation}. It would be interesting to similarly apply these fluctuation-dissipation theorems to systems undergoing varatiational dynamics. 

More interestingly perhaps are the unanswered questions as to whether or not a relativistic quantum gravity theory can be constructed from the Euclidean variational dynamics or whether variational dynamics can be used to construct a relativistic quantum gravity theory directly. Diffeomorphism invariant Osterwalder-Schrader type partial reconstructions of relativistic quantum gravity theories from Euclidean gravity theories have been proposed \cite{ashtekar1999osterwalder}\cite{ashtekar2000constructing}. Additionally, Schrader studied reflection positivity in simplicial gravity and constructed the quantum Hilbert space and observables for a particular path integral over Euclidean simplicial metrics \cite{schrader2016reflection}.   Further studies, however, may be needed to interrelate Euclidean variational dynamics and relativistic quantum gravity. Such developments may be useful in determining whether variational dynamics offers merely a new and useful tool for calculating Euclidean field configurations or whether this approach can provide further insight into the nature of quantum theory itself. 

\section*{Author Contact Information}
\noindent Brenden.McDearmon@gmail.com

\bibliographystyle{unsrt}
\bibliography{citation}
\end{multicols*}

\pagebreak
\pagenumbering{arabic}
\renewcommand*{\thepage}{A\arabic{page}}

\title{Appendix to Euclidean Quantum Gravity from Variational Dynamics}
\author{Brenden McDearmon }
\maketitle
\\
\begin{figure*}[h]
\centering
\includegraphics[width=0.85\textwidth]{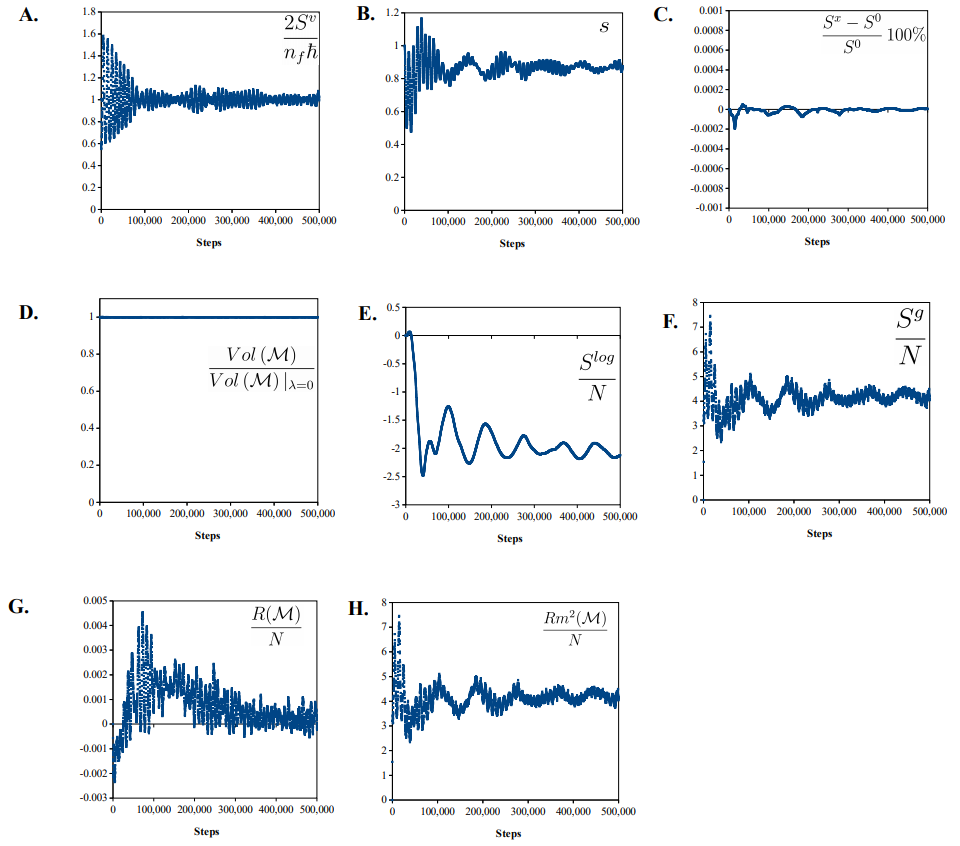}
\caption*{Figure A1: Properties of the system according to Example 4A having the action $S= s \Bigg( S^v+S^{log}+Rm^2(\mathcal{M})+0.75N\left(Vol(\mathcal{M})-Vol(\mathcal{M})\vert_{\lambda=0}\right)^2+\frac{\left(\pi^{(s)}\right)^2}{2N}+15N log(s)-S^0 \Bigg)$ where $S^v+S^{log}=\sum_{(i,j)} \left(\frac{\left(\pi^{(l)}_{ij}\right)^2}{2s^2\left(l_{ij}\right)^2} V_l -\frac{1}{2} log\left(\frac{V_l}{\left(l_{ij}\right)^4}\right) \right) $.}
\end{figure*}

\begin{figure*}[h]
\centering
\includegraphics[width=0.95\textwidth]{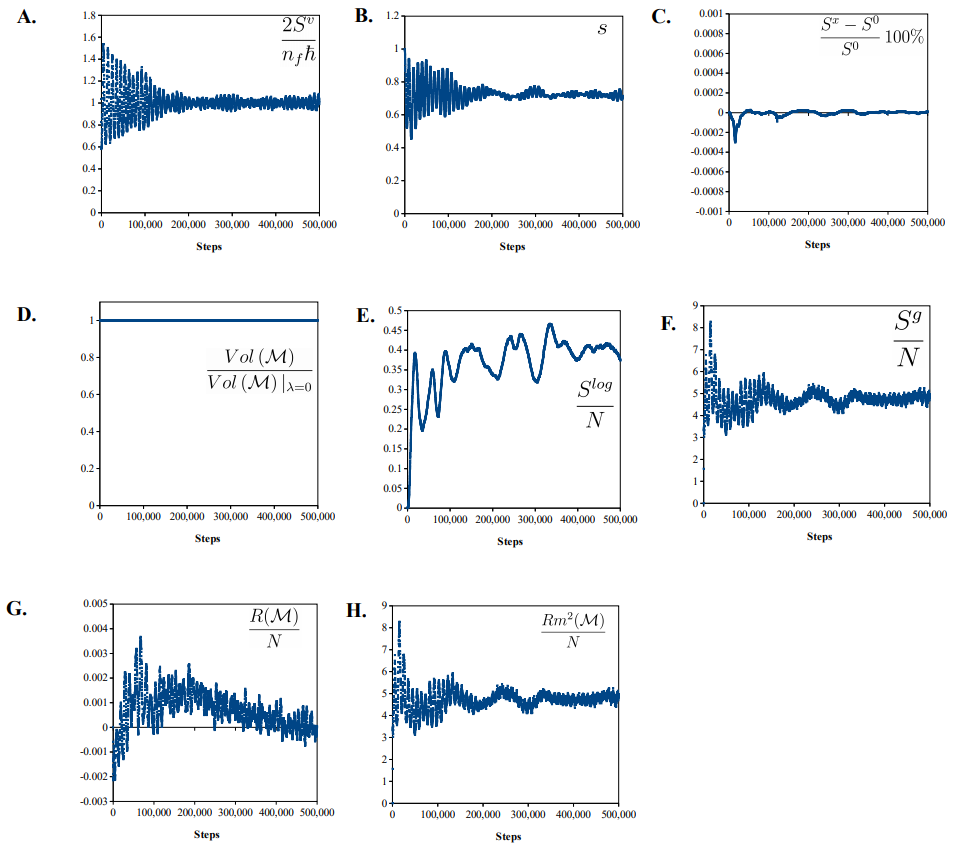}
\caption*{Figure A2:  Properties of the system according to Example 4B having the action $S= s \Bigg( S^v+S^{log}+Rm^2(\mathcal{M})+0.75N\left(Vol(\mathcal{M})-Vol(\mathcal{M})\vert_{\lambda=0}\right)^2+\frac{\left(\pi^{(s)}\right)^2}{2N}+15N log(s)-S^0 \Bigg)$ where $\,S^v+S^{log}=\sum_{(i,j)} \left(\frac{\left(\pi^{(l)}_{ij}\right)^2}{2s^2\left(l_{ij}\right)^2} w_l V_l -\frac{1}{2} log(w_l) \right) $.}
\end{figure*}

\begin{figure*}[h]
\centering
\includegraphics[width=0.95\textwidth]{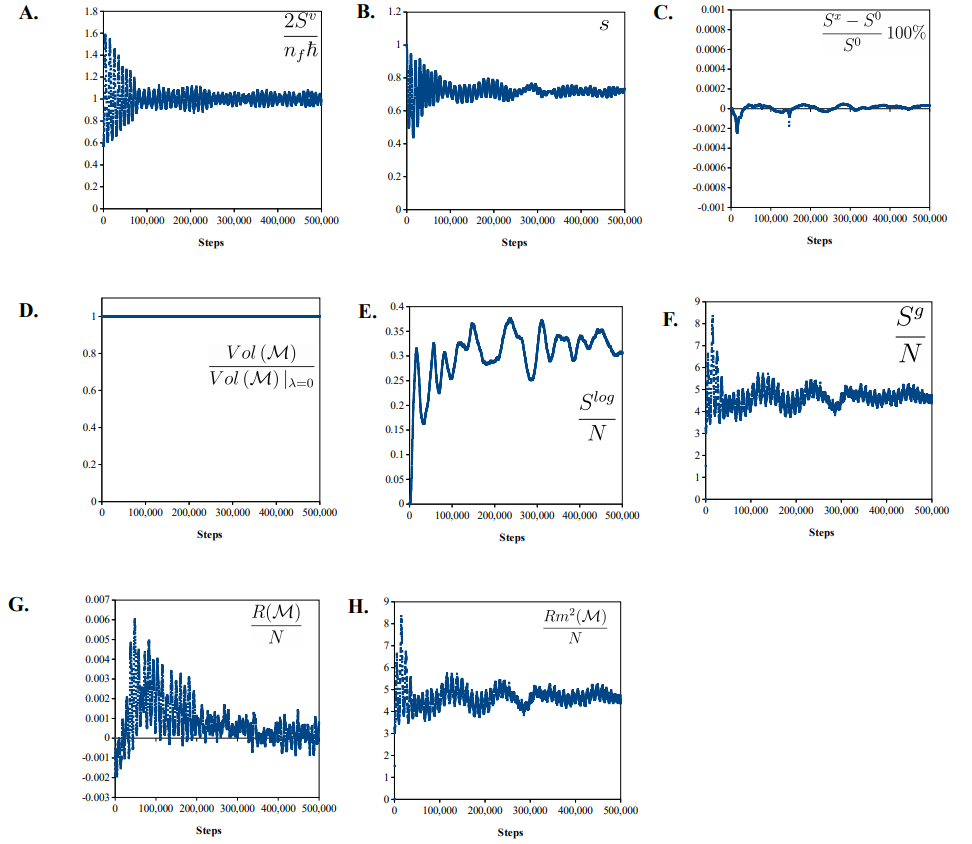}
\caption*{Figure A3:  Properties of the system according to Example 4C having the the action $S= s \Bigg( S^v+S^{log}+Rm^2(\mathcal{M})+0.75N\left(Vol(\mathcal{M})-Vol(\mathcal{M})\vert_{\lambda=0}\right)^2+\frac{\left(\pi^{(s)}\right)^2}{2N}+15N log(s)-S^0 \Bigg)$ where $\,S^v+S^{log}=\sum_{(i,j)} \left(\frac{\left(\pi^{(l)}_{ij}\right)^2}{2s^2\left(l_{ij}\right)^2} V_l -\frac{1}{2} log(w_l) \right) $. }
\end{figure*}

\end{document}